\pdfoutput=1
\pdfoutput=1
\pdfoutput=1
\pdfoutput=1
\pdfoutput=1

\documentclass[sigconf]{acmart}
\usepackage{hyperref}
\usepackage{algorithm}
\usepackage{algpseudocode}
\usepackage{amsmath}

\renewcommand\footnotetextcopyrightpermission[1]{} 
\def\BibTeX{{\rm B\kern-.05em{\sc i\kern-.025em b}\kern-.08emT\kern-.1667em\lower.7ex\hbox{E}\kern-.125emX}}
    
\usepackage{multirow}
\graphicspath{{./figures/}}
\begin{document}

%
\title{Massively Scaling Seismic Processing on Sunway TaihuLight Supercomputer}

\author{Yongmin Hu, Hailong Yang, Zhongzhi Luan and Depei Qian}
\email{{varinic, hailong.yang, 07680, depeiq}@buaa.edu.cn}
\affiliation{%
  \institution{Sino-German Joint Software Institute \\ School of Computer Science and Engineering, Beihang University, Beijing, China, 100191}
}
%

%
\renewcommand{\shortauthors}{Trovato and Tobin, et al.}

%
\begin{abstract}
Common Midpoint (CMP) and Common Reflection Surface (CRS) are widely used methods for improving the signal-to-noise ratio in the field of seismic processing. These methods are computationally intensive and require high performance computing. This paper optimizes these methods on the Sunway many-core architecture and implements large-scale seismic processing on the Sunway Taihulight supercomputer. We propose the following three optimization techniques: \textit{1)} we propose a software cache method to reduce the overhead of memory accesses, and share data among CPEs via the register communication; \textit{2)} we re-design the semblance calculation procedure to further reduce the overhead of memory accesses; \textit{3)} we propose a vectorization method to improve the performance when processing the small volume of data within short loops. The experimental results show that our implementations of CMP and CRS methods on Sunway achieve 3.50$\times$ and 3.01$\times$ speedup on average compared to the-state-of-the-art implementations on CPU. In addition, our implementation is capable to run on more than one million cores of Sunway TaihuLight with good scalability.
\end{abstract}

%
%
%

%
\keywords{Many-core Architecture, Sunway TaihuLight, Seismic Processing, Common Midpoint, Common Reflection Surface, Performance Optimization}

%

%
\maketitle

\section{Introduction}
\label{sec:introduction}
Seismic processing techniques refine seismic data to evaluate
the design of different models with cross-section images. These techniques help geologists to build models of the interested areas, which can be used to identify oil and gas reservoirs beneath the earth surface~\cite{yilmaz2001seismic}. Common Midpoint (CMP) method \cite{ottolini1984migration} and Common Reflection Surface (CRS) method \cite{jager2001common} are widely used seismic processing techniques. The general idea of the CMP method is to acquire a series of traces (\textit{gather}) that are reflected from the same mid-point under the surface. The traces are then stacked horizontally with auto-correction so that it improves the quality of the seismic data with high signal-to-noise ratio. The fold of the stack is determined by the number of traces in the CMP \textit{gather}. Different from the CMP method, the CRS method is based on the ray theory, especially the paraxial ray theory. The CRS method treats the corresponding ray at the specific reflection point in the underground as the central ray. The other rays in the neighborhood of the reflection point are regarded as the paraxial rays. All the paraxial rays determine a stacking surface. The energy stacked on the same stacking surface results in a stacked profile with high signal-to-noise ratio.

The computation demand of seismic processing is tremendous. The GPUs are commonly used as acceleration devices in seismic processing to achieve high performance. The GPUs are used to accelerate the 3D output imaging scheme (CRS-OIS) in seismic processing~\cite{ni2012gpu}, which utilizes the many-core architecture of GPU and achieves a good performance speedup on datasets with high computational intensity. The OpenCL is also used to implement the computation of semblance and traveltime~\cite{marchetti2011opencl}, that accelerates the CRS method. The existing work~\cite{lashgar2016openacc} also demonstrates the ability of OpenACC on improving the performance of seismic processing. Compared to the unoptimized OpenACC implementation, the fine-tuning technique can obtain a significant speedup. In addition to the GPU, there is also research work~\cite{marchetti2010fast} attempts to optimize seismic processing on dedicated accelerating device such as FPGA.

The Sunway TaihuLight is the first supercomputer with a peak performance of over 100 PFlops. It was ranked the first place in Top500 in June 2016. The Sunway TaihuLight uses China homemade Sunway SW26010 processor. Each Sunway processor contains four Core Groups (CGs), and each CG consists of one Management Processing Element (MPE) and 64 Computing Processing Elements (CPEs). The many-core architecture design of Sunway processor has the great potential for high-performance computing. After built in place, the Sunway processor has demonstrated its success in various scientific applications for high performance. Especially, the atmospheric dynamics~\cite{yang201610m} and earth-quake simulation~\cite{fu201718} running on the full system of Sunway TaihuLight for large-scale computation won the ACM Gordon Bell prize. Moreover, the optimization of various computation kernels, such as SpMV~\cite{liu2018towards} and stencil~\cite{ao201726}, also demonstrates the unique performance advantage of Sunway architecture. In addition to the traditional scientific applications, the Sunway processor has also shown its potential to support emerging applications. For instance, swDNN \cite{fang2017swdnn} is a highly optimized library to accelerate deep learning applications on Sunway, and swCaffe \cite{li2018swcaffe} is a deep learning framework supports large-scale training on Sunway TaihuLight.

Although existing works have explored different architectures to optimize seismic processing, it is impossible to naively adopt the existing works to Sunway due to its unique architecture design. Specifically, the following challenges need to be addressed in order to achieve good performance for the CMP and CRS methods on Sunway. First, unlike the traditional x86 processor, the design of the CPEs does not contain a cache, but a 64KB user-controlled scratch pad memory (SPM), which means without careful management, the frequent accesses to main memory could lead to severe performance degradation. Secondly, in order to achieve the ideal memory bandwidth on Sunway, the DMA transfers issued from the CPEs must contain at least 1024B data. However in the CMP and CRS methods, only a tiny data ranging from 4B to 76B is required during each computation step, which is prohibitive to achieve optimal the performance on Sunway. Moreover, the operations applied to the tiny data within short loops make it difficult to take advantage of the vector units on Sunway processor.

In order to solve the above challenges, this paper proposes a re-design of the CMP and CRS methods on Sunway processor. In addition, several optimization techniques are also proposed to adapt to the architecture features of Sunway efficiently. The experiment results demonstrate our implementation of seismic processing achieves significant speedup when scaling to massive number of Sunway cores. Specifically, this paper makes the following contributions:
\begin{itemize}
\item We propose a software cache method for seismic processing on Sunway CPEs. This method utilizes the architecture features of DMA and LDM on Sunway. When the memory access occurs, the CPE sends the data request to the software cache. After receiving the data request, the software cache retrieves data from the memory through DMA, and then send the data back to the CPE. After that, the data is buffered in the software cache to effectively alleviate the long memory access delay.
\item We re-design the Common Depth Point (CDP) procedure that dominates the performance of CMP and CRS methods to adapt to the Sunway architecture. Specifically, we combine multiple search processes onto a single CPE, and synchronize across search processes by buffering the intermediate results from each computation step. In addition, we combine the data to be accessed at each step of the search processes, and thus reduce the number of DMA accesses.
\item We propose a vectorization method to improve the computation efficiency when processing the tiny data within short loops. We first convert the global reduction operations into several independent element-wise vector operations, and then use the vector array to perform element-wise vector operations with the ending element processed separately.
\end{itemize}

The rest of this paper is organized as follows: In Section~\ref{sec:backgroud}, we introduce the background of the CMP and CRS methods, as well as the Sunway architecture. Section~\ref{sec:method} presents our design and optimization of seismic processing on Sunway to achieve massively scaling. The evaluation results are given in Section~\ref{sec:eval}. Section~\ref{sec:relatedwork} discusses the related work and Section~\ref{sec:conclusion} concludes this paper.

\section{Background}
\label{sec:backgroud}
\subsection{Common Midpoint Method}
\label{subsec:cmp}
\begin{figure}
\centering
\includegraphics[scale=0.4]{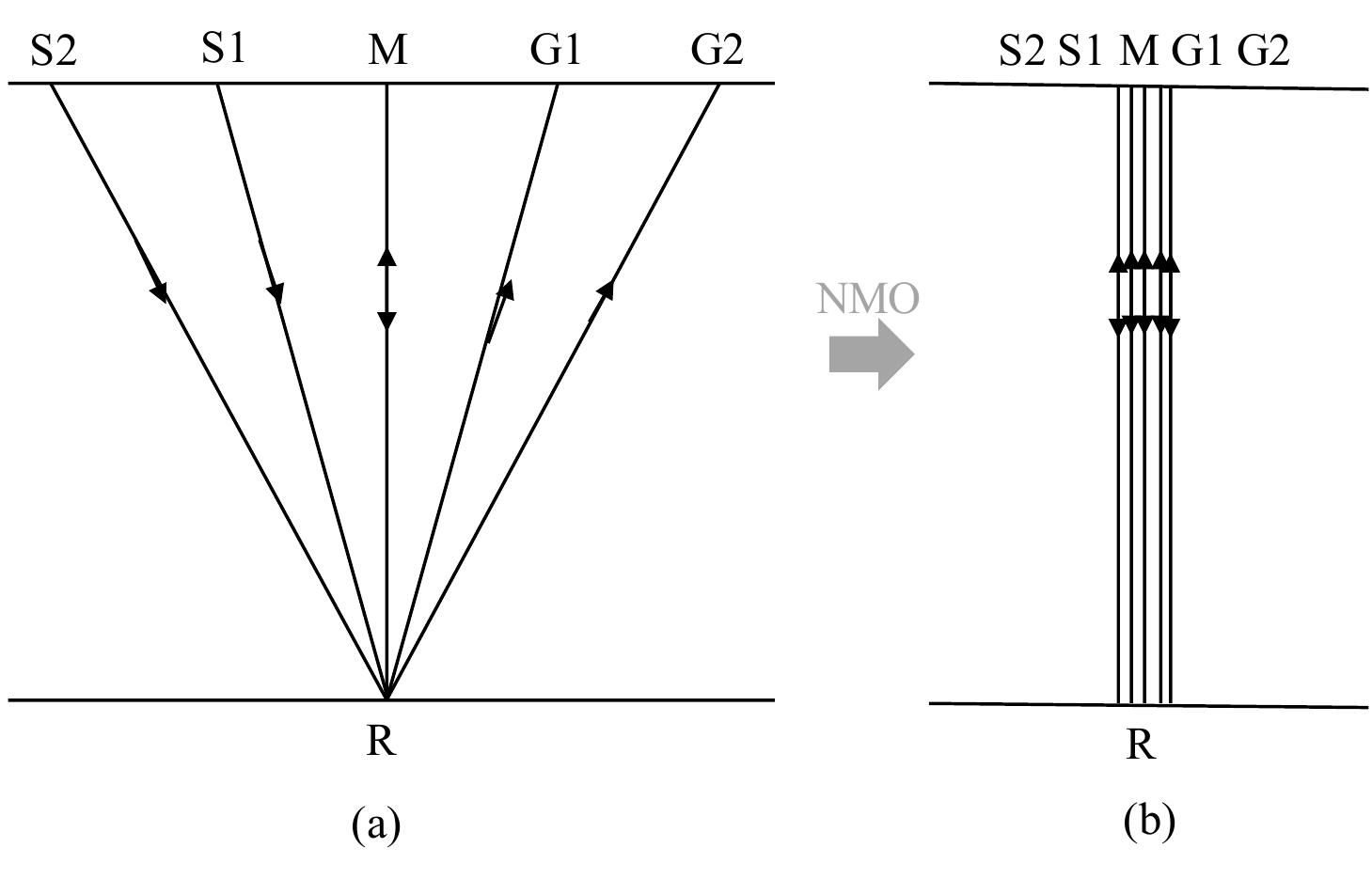}
\caption{The illustration of the common midpoint method (CMP).}
\label{fig:fig1}
\end{figure}

The fundamental idea of the CMP method is shown in Figure~\ref{fig:fig1}(a). The sound source placed at the source point $S_i$ is excited. After the sound wave is reflected by the underground reflection point $R$, the receiver on the surface receives the signal at point $G_i$. Each data record captured is a seismic trace and a group of seismic traces that share the same midpoint is called a CMP gather. When the reflection surface is horizontal and the speed does not change horizontally, the CMP gathers are equivalent to Common Depth Point (CDP) gathers. This is the seismic data this paper deals with, therefore we use the term CMP and CDP interchangeably.

In the CMP method, traces belonging to the same CMP gather are corrected and stacked, which generates a stacked trace. As shown in Figure~\ref{fig:fig1}(b), before the traces are stacked together, a Normal Moveout (NMO) correction is applied to the reflection traveltimes according to the distances between their sources and receivers, which groups signals that are produced by the same reflectors. The quality of the stacked trace depends on the quality of the NMO correction. The NMO in the CMP method is to correct the hyperbolic curve (also known as traveltime curve), which depends on the distance between the source and the receiver as well as the average velocity in which the wave propagated during the seismic data acquisition. Although the distance is known in advance, the velocity is usually unknown. Therefore, it is necessary to find the best stacking velocity.

To find the best stacking velocity, the CMP method enumerates through different velocities. For each of enumerated velocities, it computes the semblance, a coherence metric that indicates whether the traveltime curve defined by a given velocity would produce a good stacking. The semblance computation is performed over a traveltime curve that intersects seismic traces. Considering that the traces are represented by discrete samples, some points of the intersections may not align with the actual elements in the dataset. Therefore, we use the interpolation of nearby samples to estimate the seismic amplitude at that point. The Equation~\ref{eq:semblance} defines the computation for semblance. There are M traces in a single CDP, $f_{ij}$ represents the $j-th$ sample of the $i-th$ trace, and the intersection of the traveltime curve of the trace is $k$. The semblance calculation is performed in a window of length $w$, which walks through the traces of the current CDP and access $w$ samples in each intersection. The value of $w$ is determined by the sampling interval during data acquisition. In the CMP method, there is no dependency between the computation of individual CDPs, therefore they can be computed in parallel.

\begin{equation}
\label{eq:semblance}
S_c(k)=\frac{\sum_{j=k-w/2}^{k+w/2}(\sum_{i=1}^{M}f_{ij})^2}{\sum_{j=k-w/2}^{k+w/2}\sum_{i=1}^{M}f_{ij}^2}
\end{equation}

\subsection{Common Reflection Surface Method}
\label{subsec:crs}
\begin{figure}
\centering
\includegraphics[scale=0.4]{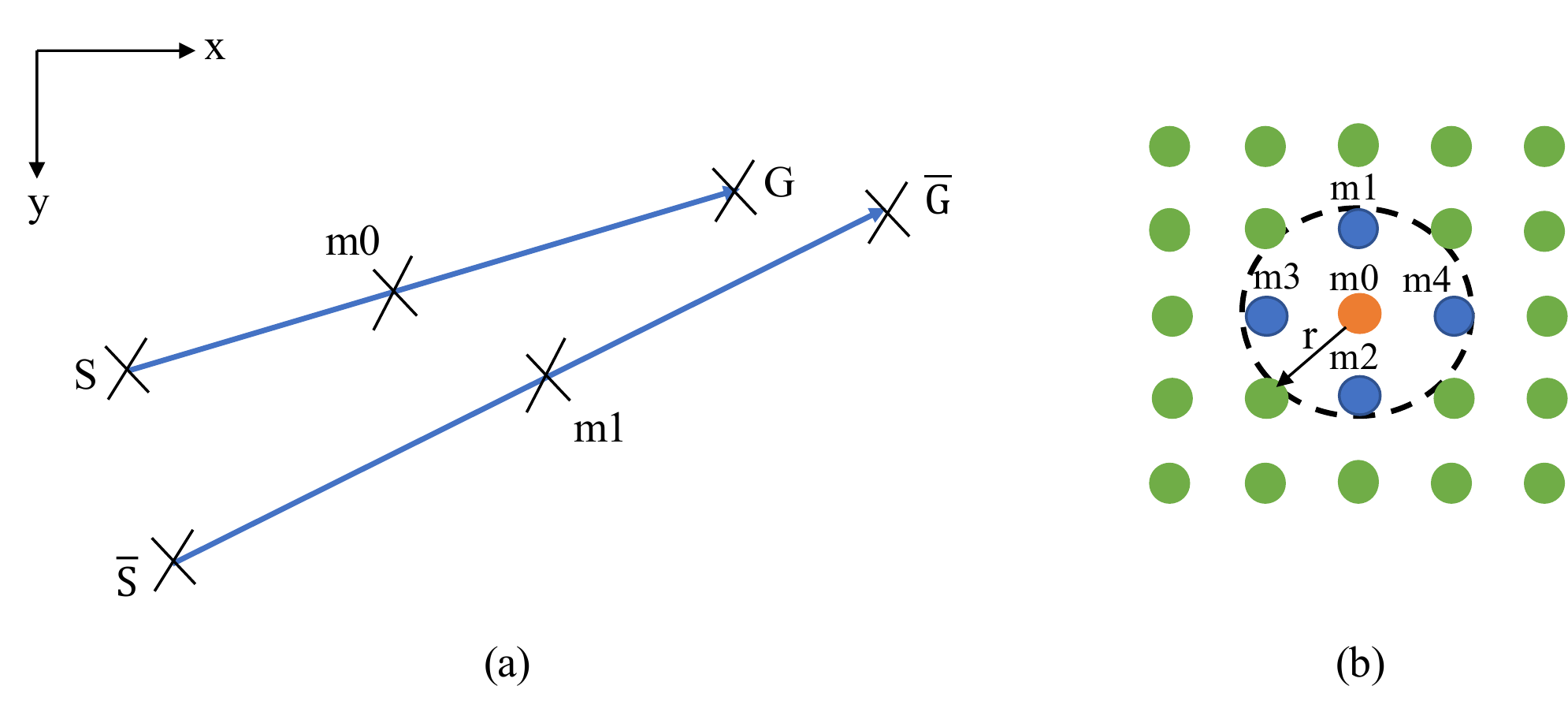}
\caption{The illustration of the common reflection surface method (CRS).}
\label{fig:fig2}
\end{figure}

As shown in Figure~\ref{fig:fig2}(a), the ray from the exciting point $S$ to the receiving point $G$ is the central ray, whereas the ray from the exciting point $\bar{S}$ to the receiving point $\bar{G}$ is the paraxial ray. The central points $m0$ and $m1$ belong to CDP0 and CDP1 respectively. According to the paraxial ray theory, when processing the central ray $SG$, it requires the data of the paraxial ray $\bar{S}\bar{G}$. Therefore, when computing the CDP0, it requires the data from CDP1. In Figure~\ref{fig:fig2}(b), the orange point $m0$ represents the central point, and the neighbors within the radius $r$ include four CDPs ($m1, m2, m3, m4$) that are represented with blue dots, and the remaining green dots are not in the neighborhood of $m0$. It also means, when processing CDP0, the data from CDP1, CDP2, CDP3 and CDP4 is required. The semblance computation in the CMP method can be easily extended to the CRS method. The only difference is to obtain the trace data of the CDPs in its neighborhood when processing the central CDP, as well as change the NMO curve to a curved surface. For large-scale processing, we partition the two-dimensional coordinates of the middle points of each CDP using grid, and map the grids to different CGs of the Sunway processor. The adjacent grids exchange data through asynchronous MPI communication.

\subsection{The Sunway Many-core Architecture}
\label{subsec:sunway}
\begin{figure}
\centering
\includegraphics[scale=0.3]{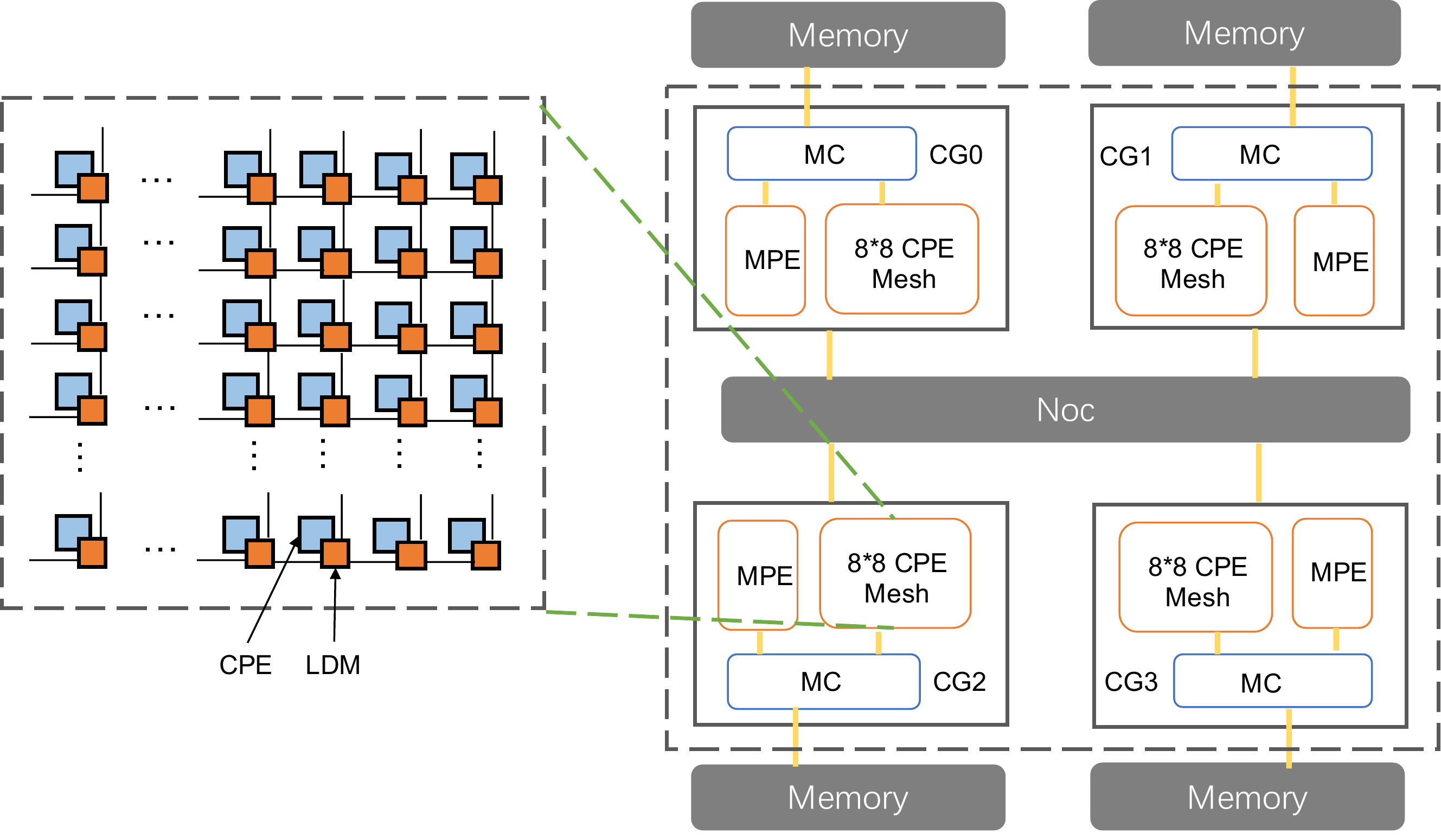}
\caption{The architecture of Sunway SW26010 processor.}
\label{fig:fig3}
\end{figure}

The Sunway TaihuLight supercomputer provides a theoretical peak performance of 125PFlops. It consists of 40,960 Sunway SW26010 processors with 1.4PB memory and an aggregated bandwidth of 4,473.16TB/s. The architecture of the Sunway processor is shown in Figure~\ref{fig:fig3}, which is composed of four core groups (CGs). Each CG contains a Management Processing Element (MPE) and 64 Computing Processing Elements (CPE), and each CG is attached of 8GB DDR3 memory. The 8GB attached memory can be accessed by both MPE and CPEs with the bandwidth of approximately 136GB/s. The MPE has 32KB L1 instruction cache and 32KB L1 data cache, in addition to 256KB L2 cache for both instruction and data. Each CPE has its own 16KB L1 instruction cache but no data cache. However, there is 64KB local device memory (LDM) on each CPE that is explicitly managed by software. The CPEs can initiate a direct memory access (DMA) operation that reads data from memory to the LDM, or writes data from the LDM to memory. The CPEs in the same row or column of the CG can communicate with each other through register communication. Each CPE has a vector unit that supports 256-bit wide vector floating-point operation. The survey paper~\cite{fu2016sunway} has shown that the memory bandwidth of Sunway processor is quite limited compared to the massive computation power. Therefore, the most effective optimization techniques on Sunway include the rational use of LDM, data transfer through register communication between CPEs, computation acceleration through vector units and data access through DMA for higher bandwidth.

\subsection{Challenges on Sunway Architecture}
\label{subsec:challenges}
As the dominant computation of the seismic processing with both CMP and CRS methods, accelerating the procedure of semblance calculation is critical to achieve satisfactory performance on Sunway. However, the data access pattern during the semblance calculation is prohibitive to obtain good performance on Sunway for two reasons: \textit{1)} the data accesses are random, which leads to high memory access latency due to the lack of data cache on CPEs; \textit{2)} the volume of data accesses is quite small, which is unable to fully utilize the precious DMA bandwidth as well as the vector units for performance acceleration. The specific challenges are as follows:


\begin{itemize}
\item The random data accesses during the semblance calculation deteriorate the performance of seismic processing on Sunway. Due to the lack of data cache on CPEs, a software cache method is necessary to buffer the data accessed from main memory by using the limited LDM on CPEs effectively.
\item The semblance calculation only accesses a small volume of data, which is hard to fully utilize the DMA bandwidth. Therefore, it is necessary to re-design the process of semblance calculation by combining the computations and buffering the intermediate results on each CPE in order to improve bandwidth utilization. 
\item In addition to the low bandwidth utilization, the small volume of data during the semblance calculation also prohibits the exploration of vectorization. Therefore, to utilize the vector units on Sunway, it requires revealing the vectorization potential by adjusting the computation pattern of semblance calculation.
\end{itemize}
\section{Re-designing the Seismic Processing for Massively Scaling}
\label{sec:method}

\subsection{Design Overview}
\label{subsec:overview}
\begin{figure*}
\centering
\includegraphics[scale=0.4]{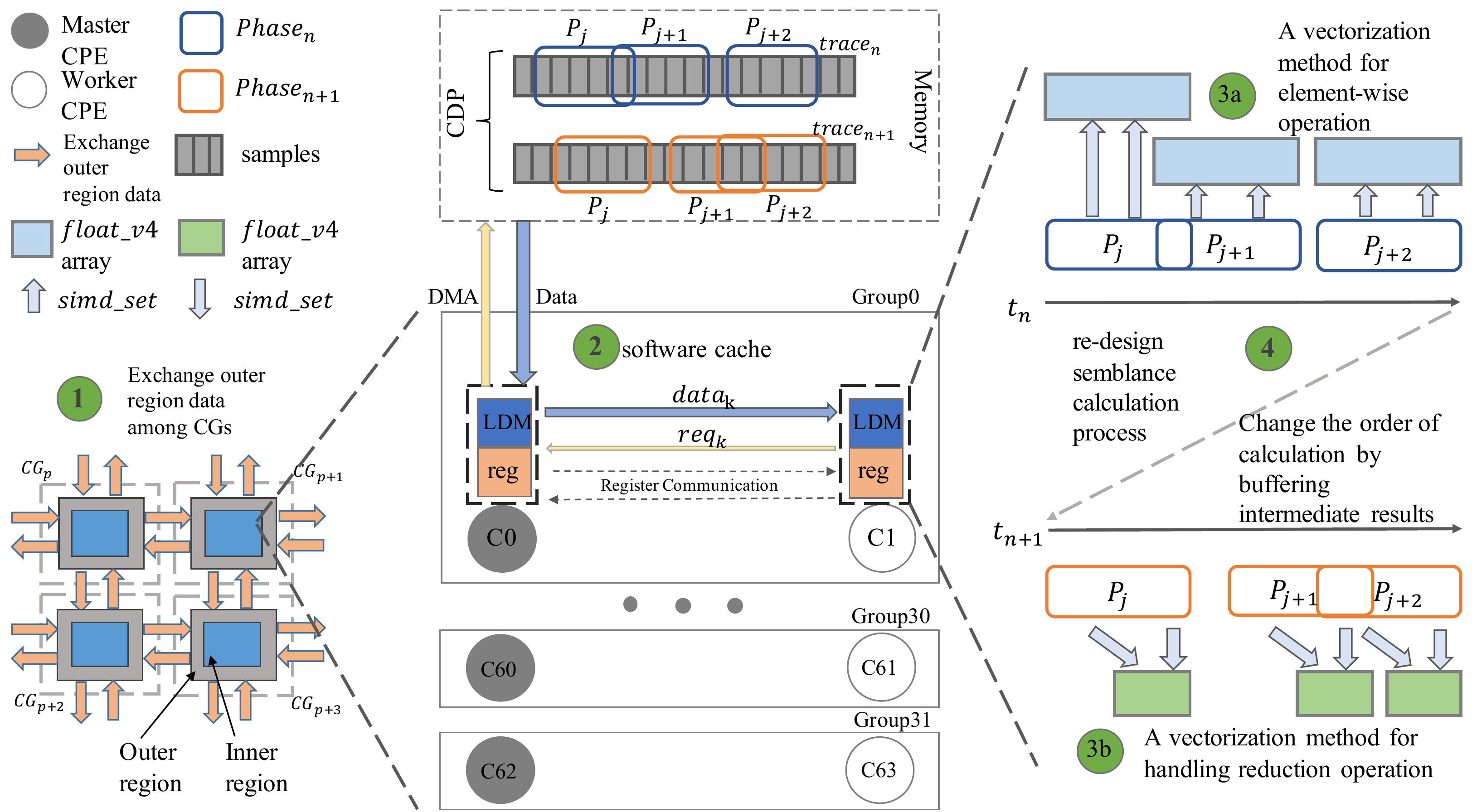}
\caption{The design overview of CDP computation on Sunway architecture.}
\label{fig:fig4}
\end{figure*}

Figure~\ref{fig:fig4} shows the design and optimization of the CDP computation of the CMP and CRS methods on Sunway architecture. Firstly, the MPE on each CG reads the partitioned seismic data. Seismic data consists of several CDPs, and each CDP contains several traces, each of which is composed of $ns$ samples. For the CMP method, the computation on each CDP is independent from the rest. Whereas for the CRS method, the computation of the central CDP requires data from the surrounding CDPs. In such case, the MPE calculates the two-dimensional coordinates of the \textit{middle point} of each CDP, and then divides the inner and outer regions according to the two-dimensional coordinates. The calculation of the outer region involves the region of the adjacent mesh, which requires the CDPs in the outer region to be sent to the adjacent mesh. As shown in Figure~\ref{fig:fig4} step \textcircled{1}, the data transfer of the outer region and the calculation of the inner region is performed simultaneously through asynchronous MPI. After the central CDP receives the data from surrounding CDPs, the CDP computation is the same in both CRS and CMP methods. Therefore, we take the computation of a single CDP for an example to illustrate the optimizations we have applied on each Sunway CG.

The CDP computation involves the semblance calculation that walks through the traces of the current CDP. In order to improve the performance of the CDP computation on Sunway, we propose several techniques to re-design and optimize the computation procedures. First, we use the master CPE and worker CPEs collaboratively to implement a software cache to eliminate random data accesses at the intersection (step \textcircled{2}). Second, we propose a vectorization method to improve the computation efficiency when processing the tiny data within short loops (step \textcircled{3}). Third, we re-design the calculation process so that each worker CPE can process multiple sample-NMO velocity pairs simultaneously to further improve bandwidth utilization (step \textcircled{4}).

To better illustrate how our proposed techniques work together, we take the processing of one CDP as shown in the upper part of Figure~\ref{fig:fig4} for example. There are two adjacent traces ($trace_{n}$ and $trace_{n+1}$) in a single CDP stored in continuous memory region, and each trace contains three (sample, NMO velocity) pairs (e.g., $P_{j}$, $P_{j+1}$, $P_{j+2}$). When memory access occurs, the software cache first takes into effect (step \textcircled{2}.) Every two adjacent CPEs in each row of the CPE mesh are organized into a group, with one of them serving as the master CPE and the other one serving as the worker CPE. The worker CPE first sends a data request to the master CPE in its own group through register communication. After the master CPE receives the request, it retrieves the data from the memory through DMA, then sends the requested data back to the worker CPE. The requested data is buffered in the LDM of the master CPE. The vectorization method (step \textcircled{3a} and \textcircled{3b}) converts the reduction operation into independent element-wise vector operations, then uses the vector array to perform element-wise vector operation with the ending elements processed separately. The re-designed calculation process (step \textcircled{4}) synchronizes the processing of $P_{j}$, $P_{j+1}$ and $P_{j+2}$ in sequence on $trace_{n}$, and buffers their intermediate results. Then, the CPE group continues to process next trace ($trace_{n+1}$). The above steps are repeated until the last trace is processed. After all the (sample, NMO velocity) pairs have been processed, the computation of a single CDP completes.

\subsection{Improving Parallelism within a CG}
\label{subsec:singlecg}
\begin{figure}
\centering
\includegraphics[scale=0.26]{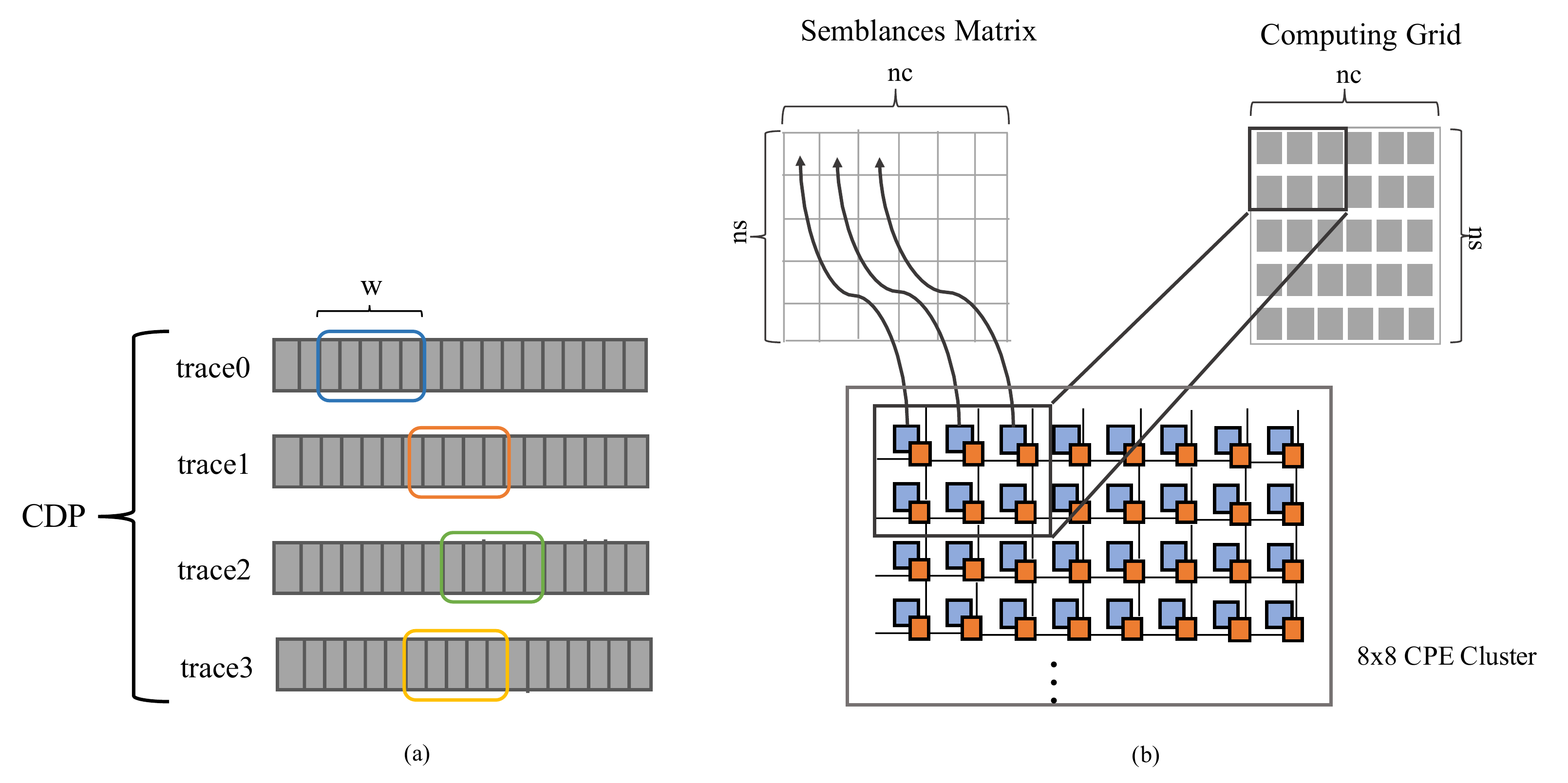}
\caption{The data structure of CDP (a), and the mapping of (sample, NMO velocity) pair search to CPEs (b).}
\label{fig:fig5}
\end{figure}

Since the CDP computation dominates the execution time of seismic processing, the optimization of the CDP on a CG is critical to fully exploit the performance of Sunway processor. The maximal number of traces in a CDP is the $fold$ of the dataset and the total number of CDPs in a dataset is denoted as $ncdps$. Each seismic trace is represented by an array, where each element is a sample. We assume that the seismic traces have the same number of samples ($ns$) across all CDPs, which is widely accepted in literature~\cite{gimenes2018evaluating}. Figure~\ref{fig:fig5}(a) shows that a CDP contains 4 traces, each of which contains several samples. In the same CDP, two adjacent traces are stored in continuous memory region. In addition, the samples of a single trace are also stored continuously. As shown in Figure~\ref{fig:fig5}(a), the center of the four colored boxes is the intersection of the traveltime curve with the four traces. The semblance is computed within a window of width $w$, which also represents the number of samples in each color boxes.

In order to achieve parallel processing of the CDP on a single CG, we chose a grid-based search method. The entire computation is divided into three phases, including initialization, calculation and result writing back. The initialization phase is performed on the MPE. Firstly, the CDP data is accessed, and the NMO velocity array is generated according to the upper and lower bounds of the NMO velocity. Then the halfpoints are computed that are necessary for calculating the traveltime curve. The NMO velocity is stored in an array of size $nc$, whereas the halfpoints are stored in another array, each element of which corresponds to a trace in the CDP. The MPE then creates a semblance matrix $S$ with size of $ns \times nc$ in memory, which is used in the semblance computation to find the most coherent NMO speed. At the same time, the MPE also creates an array with size of $ns$ in memory to store the most coherent NMO velocity.

The calculation phase involves finding the most coherent NMO velocity for each sample from the NMO velocity array. For each sample, we enumerate the elements in the NMO velocity array, and compute the semblance of each (sample, NMO velocity) pair by walking through the traces of the current CDP according to the traveltime curve. Each (sample, NMO velocity) pair is independent from each other and can be processed in parallel. Figure~\ref{fig:fig5}(b) shows how the computing grid is divided among the CPEs and how the results are written back to the semblance matrix. Each point in the computing grid represents a (sample, NMO velocity) pair. In this example, there are 6 NMO speeds ($nc$) and 5 samples ($ns$). The enlarged area shows how the points in the computing grid are mapped to CPEs, which means each CPE is responsible for computing a (sample, NMO velocity) pair. After filling in the entire semblance matrix, we need to find the NMO velocity with biggest semblance value for each sample. This velocity is the best coherent velocity required.

The computation procedure of a single CDP is shown in Algorithm~\ref{code:cdp}. For each CDP, it enumerates the sample-NMO velocity pairs (line~\ref{code:eachpair}), and then finds the intersection of the traveltime curve and traces. At each intersection, it first obtains the halfpoint of the current trace (line~\ref{code:tracestart}-\ref{code:traceend}), then accesses the data with size of $w$ (line~\ref{code:datawstart}-\ref{code:datawend}), and finally retrieves the data computed in a window of width $w$ (line~\ref{code:windowstart}-\ref{code:windowend}). Each trace has its own corresponding halfpoints, therefore the accesses to halfpoints are continuous when walking through the traces sequentially. Based on this observation, we can reserve a space $h\_s$ of size $size\_h$ on LDM to prefetch the halfpoints in advance and save them in $h\_s$. We can control the amount of prefetched data in LDM by adjusting $size\_h$.

\begin{algorithm}[htpb]
\scriptsize
\caption{The computation of a single CDP on a CG}
\label{code:cdp}
\begin{algorithmic}[1]
\Function {CDP}{sample-NMO pairs}
	\For{each sample-NMO pair $i$}\label{code:eachpair}
		\For{$j = 0 \to w$}
			\State $\_num[j] \gets 0$
		\EndFor
		\State $\_ac\_linear \gets 0$
		\State $\_den \gets 0$
		\For{$t = 0 \to ntrace$}
			\If{$t \% size\_h == 0$}\label{code:tracestart}
			\State{prefetch $size\_h$ halfpoints through DMA}
			\EndIf\label{code:traceend}
			\State{calculate index $k1$ of random data access}\label{code:datawstart}
			\State {get data of size $w$ by $k1$ through DMA}\label{code:datawend}
 			\For{$j = 0 \to w$}\label{code:windowstart}
				\State $v \gets (cache[j+1]-cache[j])*x+cache[j]$
				\State $\_num[j] \gets \_num[j] + v$
				\State $\_den \gets \_den + v*v$
				\State $\_ac\_linear \gets \_ac\_linear + v$
			\EndFor\label{code:windowend}
		\EndFor
	\EndFor
\EndFunction
\end{algorithmic}
\end{algorithm}

%

\subsection{Eliminating Random Memory Access}
\label{subsec:randommemory}

\subsubsection{Software Cache within a CG}
\label{subsubsec:cache}

Due to the limited memory bandwidth on Sunway, we propose a software cache to alleviate the long memory access delay caused by random data access. We design a software cache, that is, two adjacent CPEs in each row of the CPE mesh are organized into a group, and one CPE in each group is selected to act as the software cache of the group. The selected CPE for caching is the master CPE, and the other one is the worker CPE. When memory access occurs, the master CPE accesses the data through DMA and distributes the data to the worker CPE through register communication. Existing research~\cite{xu2017benchmarking} reveals that when the accumulative data size of the DMA accesses from the 64 CPEs within a CG is less than 1024B, the achievable DMA bandwidth is proportional to the size of data accesses. In both CMP and CRS methods, the maximum size of data access is 76B. Therefore, the proposed software cache is capable to combine multiple DMA accesses, which not only reduces the number of memory accesses, but also increases the achievable DMA bandwidth.



\begin{figure}
\centering
\includegraphics[scale=0.45]{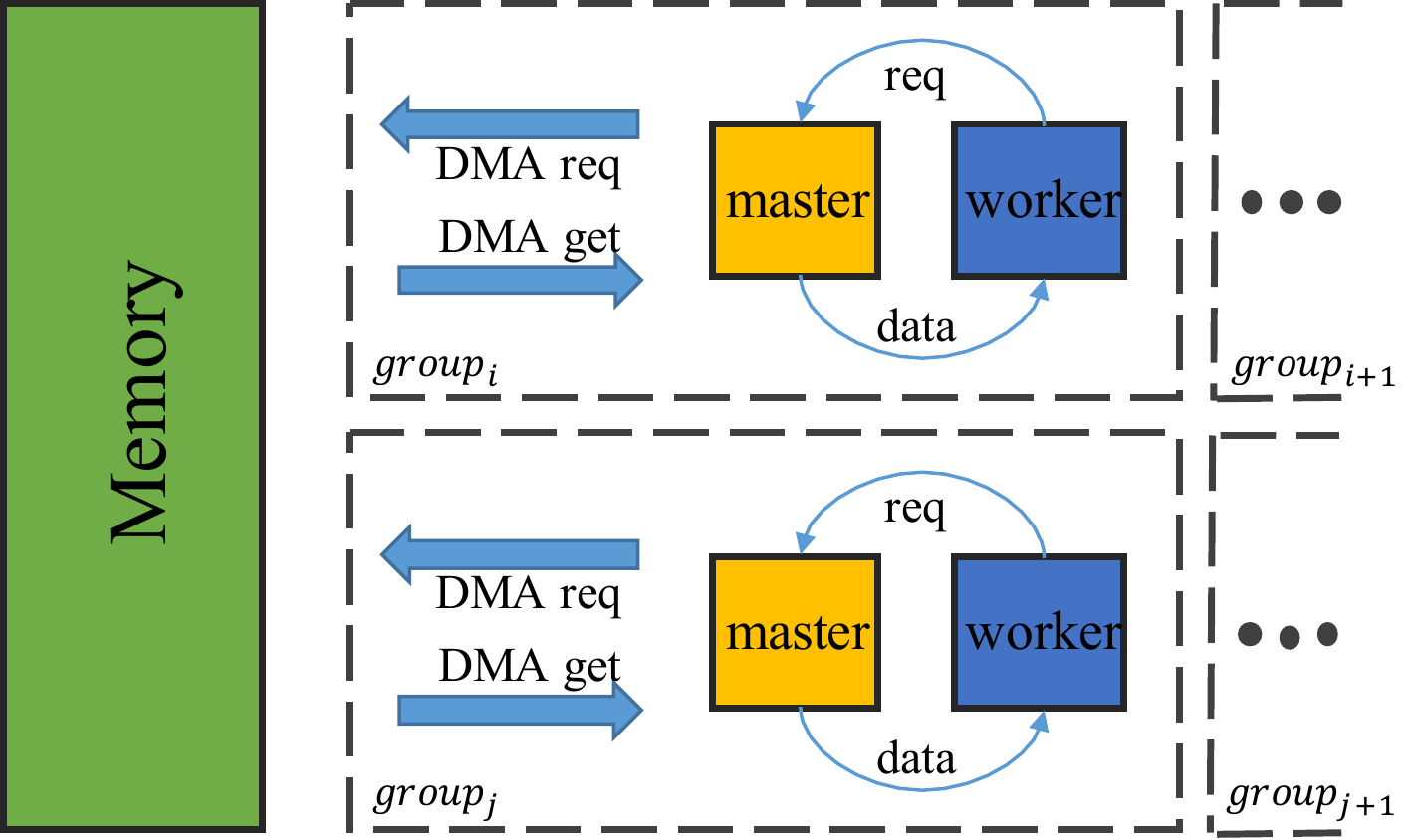}
\caption{The memory accesses at the software cache for the CDP computation.}
\label{fig:fig6}
\end{figure}

When designing the software cache, we also consider the computation characteristics of CDP. The memory accesses at the software cache during the CDP computation are shown in Figure~\ref{fig:fig6}. We denote the processing of a trace as a phase. The master and worker CPEs calculate their corresponding memory region of data access, and the worker CPE sends the requested memory region to the master CPE. After the master CPE receives the request, it identifies the minimum and maximum memory address among the regions, and then copies the data between the minimum and the maximum address to the LDM of the master CPE. The master CPE sends back the data to the worker CPE based on the requested memory region. Then both the master and worker CPEs start their corresponding calculations. When the master and worker CPEs finish processing current trace, they proceed to the next trace.

\begin{figure}
\centering
\includegraphics[scale=0.45]{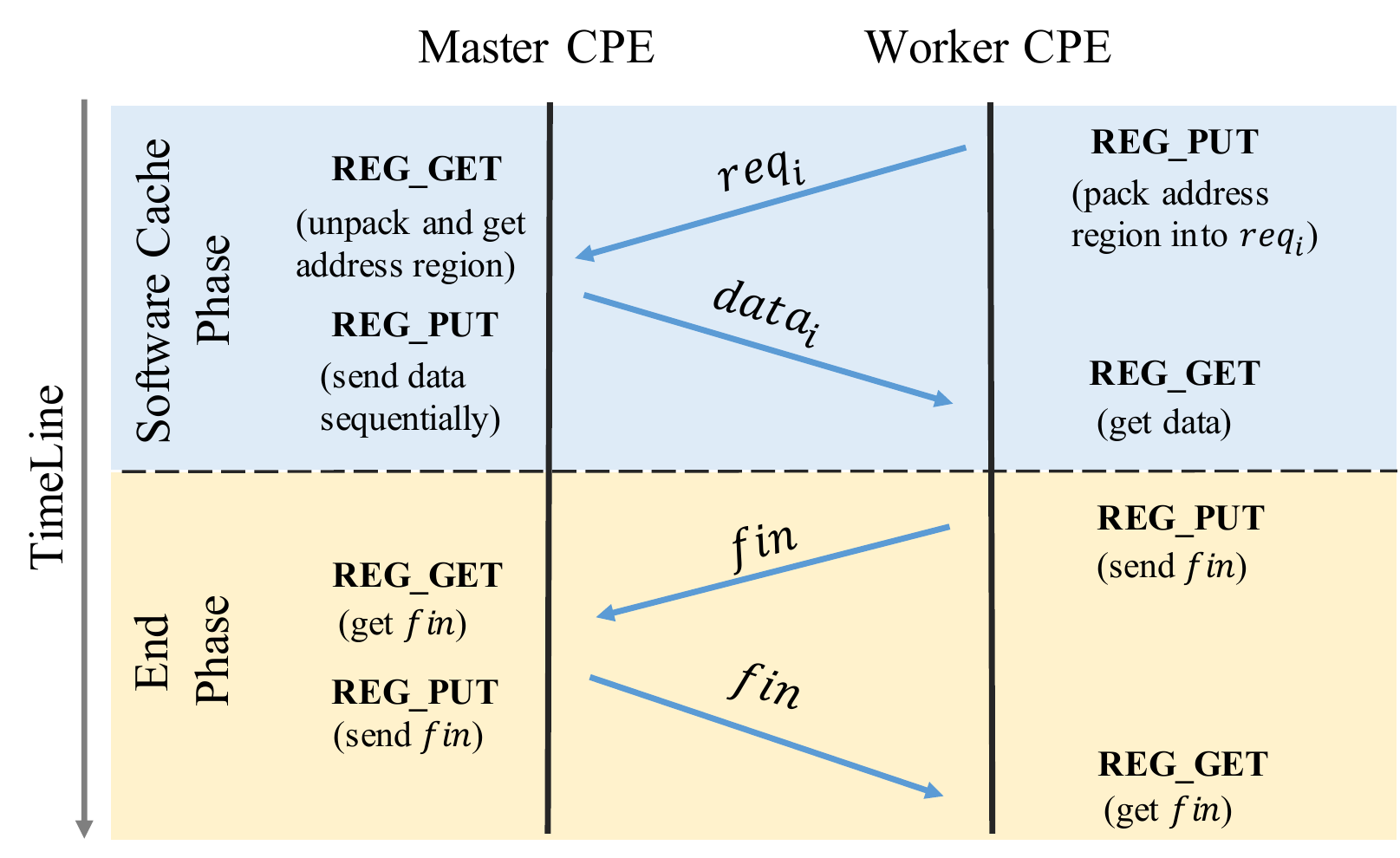}
\caption{The synchronization-free mechanism for the communication between the master and worker CPEs.}
\label{fig:fig7}
\end{figure}

We implement a synchronization-free mechanism to reduce synchronization overhead for the communication between the master and worker CPEs. As shown in Figure~\ref{fig:fig7}, a request signal is sent to the master CPE from the worker CPE. After obtaining the data from the memory, the master CPE sends the data back to the worker CPE. We name the above procedure as a round of communication. After multiple rounds of communication, the worker CPE finishes the calculation and sends a $fin$ signal to the master CPE. After the master CPE receives the $fin$ signal, it also sends a $fin$ signal to the worker CPE, and releases the software cache. Then the master CPE enters the calculation mode to complete the remaining calculations. Since we assign more tasks to the master CPE than the worker CPE, the master CPE always completes its calculation later than the worker CPE. It is clear that with the above mechanism, no synchronization is required for the communication between the master and worker CPEs.

\subsubsection{Re-designing the Computation for Semblance}
\label{subsubsec:semblance}
\begin{figure}
\centering
\includegraphics[scale=0.35]{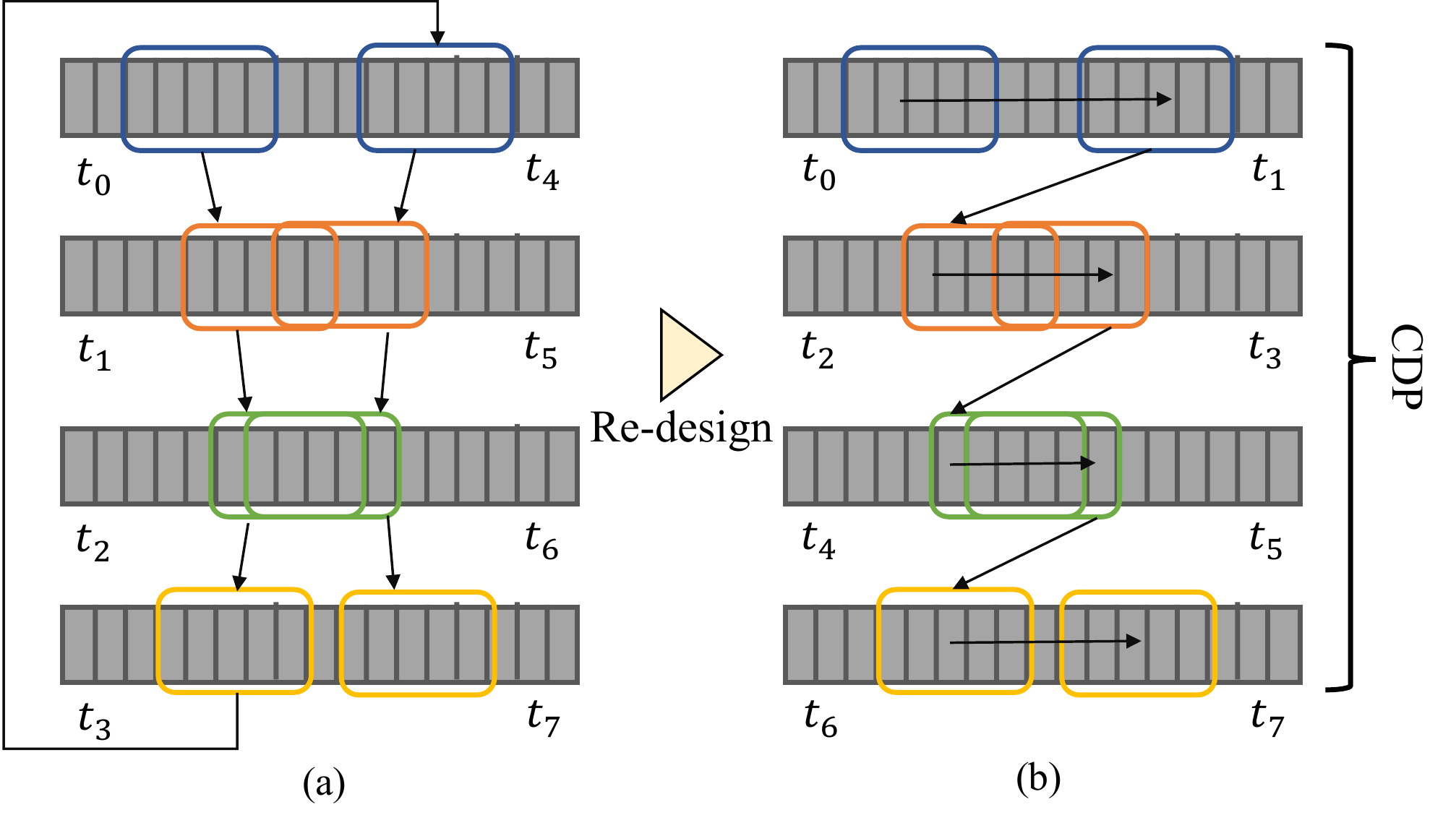}
\caption{The original (a) and re-designed (b) computation procedure of semblance calculation.}
\label{fig:fig8}
\end{figure}

Due to the poor data locality in the semblance calculation, we re-design the semblance calculation procedure and the data access pattern in order to reduce the number of memory accesses and improve the data reuse. The original procedure of the semblance calculation is shown in Figure~\ref{fig:fig8}(a). For each trace, the computing grid contains $ns \times nc$ points. To perform the search, there are $ns \times nc$ intersections randomly distributed in the trace, which leads to a large number of random memory accesses. As shown in Figure~\ref{fig:fig8}(b), after the calculation re-design, we process all the intersections in a trace continuously, and save the intermediate results from the computation of each intersection before moving on to the next trace. The above procedure is repeated until the last trace is processed. After the re-design, the calculation of next trace can reuse the intermediate results from the previous trace in the LDM. In addition to the calculation re-design, we also re-design the data access pattern. Each CPE has $ns \times nc \div cores$ intersections of a trace. Before the re-design, each data access happens in a different time period, with no opportunity to merge data accesses or reuse the data. However, after the re-design, the intersections on each CPE are processed continuously. Based on this property, we identify the minimum ($min\_la$) and maximum ($max\_lb$) memory regions for all the samples within a trace, and prefetch the data between the memory region $min\_la$ and $max\_lb$ before processing the trace.

Algorithm~\ref{code:redesign} presents the re-designed procedure of the semblance calculation. Multiple sample-NMO velocity pairs are processed simultaneously on a single CPE. For each sample-NMO velocity pair, the $\_num$, $\_ac\_linear$ and $\_den$ variables used during the computation have been expanded with one more dimension respectively for buffering data (line~\ref{code:initstart}-~\ref{code:initend}), compared to Algorithm~\ref{code:cdp}. After initialization, the traces in a CDP are processed in sequence (line~\ref{code:eachtrace}) and the data \textit{halfpoints} is prefetched before a new trace is processed (line~\ref{code:halfstart}-~\ref{code:halfend}). For the current trace, the memory addresses of the data accesses are calculated for each sample-NMO velocity pair and kept in the $k1$ array (line~\ref{code:k1start}-~\ref{code:k1end}). Then, the maximum and minimum memory address in $k1$ array is identified (line~\ref{code:minmaxstart}-~\ref{code:minmaxend}) and used to determine the memory range ($length$) of data accesses (line~\ref{code:length}). The data within the memory range is copied to LDM at one time through DMA operation (line~\ref{code:cachestart}). Finally, the calculation is performed in a window size of $w$ for each sample-NMO velocity pair and the intermediate results are kept in the $\_num$, $\_ac\_linear$ and $\_den$ arrays (line~\ref{code:window2start}-~\ref{code:window2end}). After processing the current trace, the algorithm continues to process the next trace until all traces in the CDP are processed. The final results are stored in the $\_num$, $\_ac\_linear$ and $\_den$ arrays.


Compared to the CMP method, the CRS method is more computationally intensive. Due to the limited LDM on each CPE, we cannot prefetch the data of all merged intersections at once. Therefore it is necessary to tile the merged intersections in order to assign the computation to multiple tasks. For instance, if the original loop size is $len\_i$ to process the merged intersections. After tiling, the loop is divided into two tightly nested loops. The inner loop size is $tile\_size$ and the outer loop size is $len\_i \div tile\_size$. The LDM space occupied by the merged intersections is proportional to the $tile\_size$ other than the $len\_i$. Therefore, the tile operation allows the program to effectively control the usage of LDM by the merged intersections .

\begin{algorithm}[htpb]
\scriptsize
\caption{Re-designing the computation of semblance}
\label{code:redesign}
\begin{algorithmic}[2]
\Function {re-design}{sample-NMO pairs}
	\For{each sample-NMO pair $i$}\label{code:initstart}
		\For{$j = 0 \to w$}
			\State $\_num[i][j] \gets 0$
		\EndFor
		\State $\_ac\_linear[i] \gets 0$
		\State $\_den[i] \gets 0$
	\EndFor\label{code:initend}
	\For{$t = 0 \to ntrace$}\label{code:eachtrace}
		\If{$t \% size\_h == 0$}\label{code:halfstart}
		\State{prefetch $size\_h$ halfpoints through DMA}
		\EndIf\label{code:halfend}
		\For{each sample-NMO pair $i$}\label{code:k1start}
			\State{calculate index $k1[i]$ of random data access}
		\EndFor\label{code:k1end}
		\For{each sample-NMO pair $i$}\label{code:minmaxstart}
			\State{find the min and max val $min\_la$ $max\_lb$ in $k1[i]$}
		\EndFor\label{code:minmaxend}
		\State $length \gets max\_lb-min\_la$\label{code:length}
		\State {get data of size $length$ by $min\_la$ through DMA}\label{code:cachestart}
		\For{each sample-NMO pair $i$}\label{code:window2start}
 			\State $k \gets k1[i]-min\_la$
			\For{$j = 0 \to w$}\label{code:shortloopstart}
				\State $v \gets (cache[k+j+1]-cache[k+j])*x[i]+cache[k+j]$
				\State $\_num[i][j] \gets \_num[i][j] + v$
				\State $\_den[i] \gets \_den[i] + v*v$
				\State $\_ac\_linear[i] \gets \_ac\_linear[i] + v$
			\EndFor\label{code:shortloopend}
		\EndFor\label{code:window2end}
	\EndFor
\EndFunction
\end{algorithmic}
\end{algorithm}

\subsection{Exploiting Vectorization}
\label{subsec:vectorization}

\begin{figure}
\centering
\includegraphics[scale=0.45]{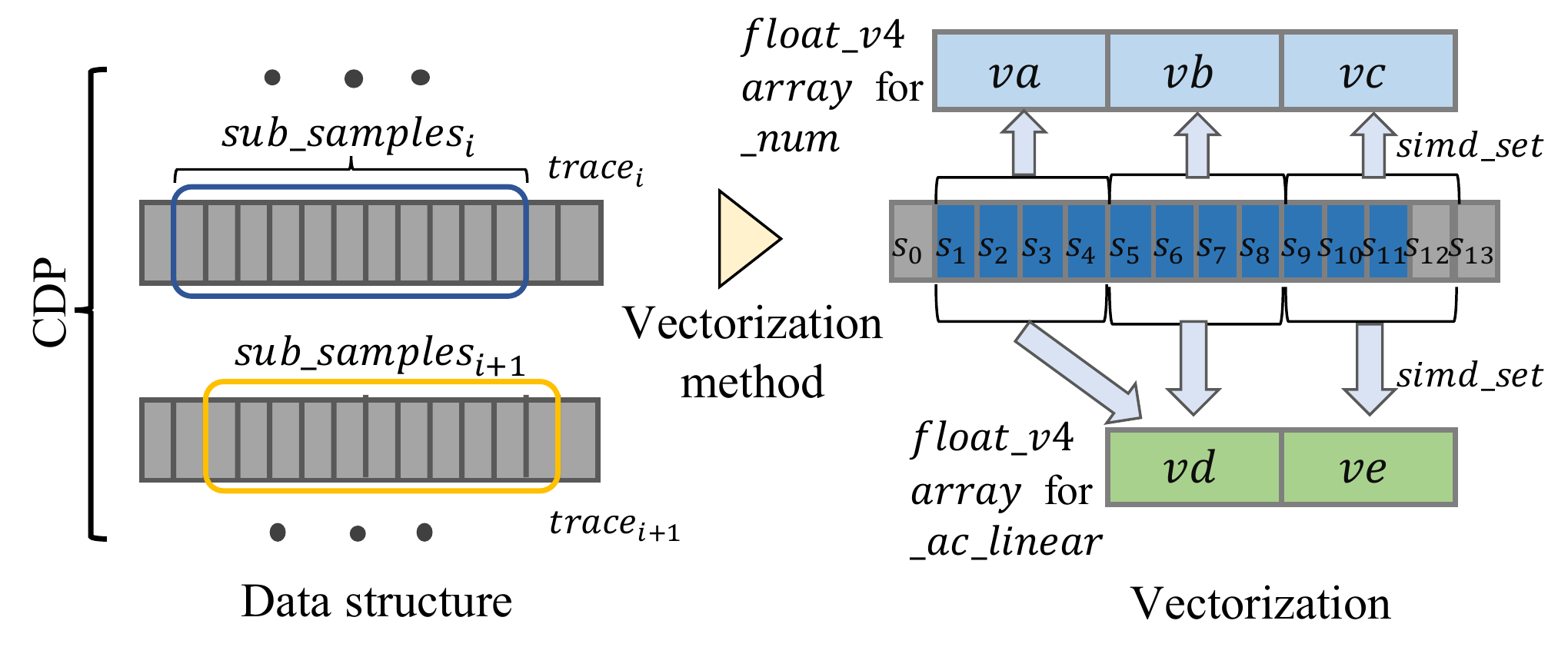}
\caption{An illustrative example for applying the vectorization method on a single CDP.}
\label{fig:fig9}
\end{figure}

We further exploit the opportunity for vectorization after the re-design of semblance calculation. As shown in Algorithm~\ref{code:cdp}, each sample-NMO velocity pair maintains the corresponding $\_num$ array and $\_ac\_linear$, $\_den$ variables. In the innermost loop, the element-wise vector calculations are applied to the $\_num$ array, whereas the reduction calculations are applied to $\_ac\_linear$ and $\_den$ variables. As shown in Figure~\ref{fig:fig9}, the random accessed data is only a small portion of the samples from each trace, which is recorded as $sub\_samples$. To vectorize the above calculations on Sunway, two challenges need to be addressed. Firstly, the window size $w$ may not be a multiple of four, which is the width of the vector unit on Sunway. Considering the reduction operation, if we directly vectorize the innermost loop, then for each $sub\_samples$, the ending data cannot be effectively vectorized which requires additional processing. In particular, if $w$ is small, which means the innermost loop is a short loop, then the overhead of processing the ending data outweighs the benefit of vectorization. Secondly, $sub\_samples$ may not be 32B aligned in LDM due to the random data access. On Sunway, the unaligned SIMD load/store throws an exception and then is split into several normal load/store instructions, which fails to exploit the computation capability of the vector unit.

Figure~\ref{fig:fig9} shows an example on how the vectorization method is applied to a single CDP. In order to load the unaligned $sub\_samples$ into the vector register, we use the $simd\_set\_floatv4$ instruction that can load four unrelated float variables into the $float\_v4$ variable, without requiring these four variables to be 32B aligned. However, compared to the standard $simd\_load$ instruction, it requires multiple LDM accesses. For element-wise vector operations, we use the vector array that consists of the $float\_v4$ vector variables with the length of $ceil(\frac{w \times1.0}{4})$. Taking Figure~\ref{fig:fig9} as an example, when $w$ equals to 11, the $sub\_samples_i$ contains 11 elements of $s_1 \sim s_{11}$, and the $\_num$ vector array contains three $float\_v4$ variables of $va$, $vb$, and $vc$, altogether representing the $sub\_samples_i$. Three $simd\_set\_floatv4$ instructions are required to load $s_1 \sim s_{12}$ into the vector array in order to perform element-wise vector calculations. For the reduction calculation, the vector array consists of two $float\_v4$ variables $vd$ and $ve$. The $s_1\sim s_4$ and $s_5\sim s_8$ are first loaded into $vd$ sequentially, and then the vector calculation is performed. After that, the $s_9 \sim s_{12}$ are loaded into $ve$ for the vector calculation. When the calculation of the current trace ($trace_i$) completes, it proceeds to the next trace ($trace_{i+1}$a). After all the traces in a CDP are processed, the $\_num$ vector array contains the results of element-wise vector operations on all $sub\_samples$ of the CDP. To derive the results of reduction calculations, the four elements in $vd$ and the first three elements in $ve$ need to be accumulated.

In Figure~\ref{fig:fig9}, the data in $s_{12}$ is invalid, and thus the result in this corresponding position is also invalid for both element-wise vector calculation and reduction calculation. Although it seems to consume extra space and computing resources, such design can effectively reduce the overhead of processing the data at the end of $sub\_samples$ in the short loop. With the re-design of semblance calculation, the intermediate results of processing multiple sample-NMO search pairs need to be buffered on the same CPE. The intermediate results of element-wise vector calculations and reduction calculations including vector arrays $\_num$, $\_ac\_linear$ and $\_den$ are also need to be buffered.

\subsection{Asynchronous Parallel Processing among CGs}
\label{subsec:asynchronous}
\begin{figure}
\centering
\includegraphics[scale=0.45]{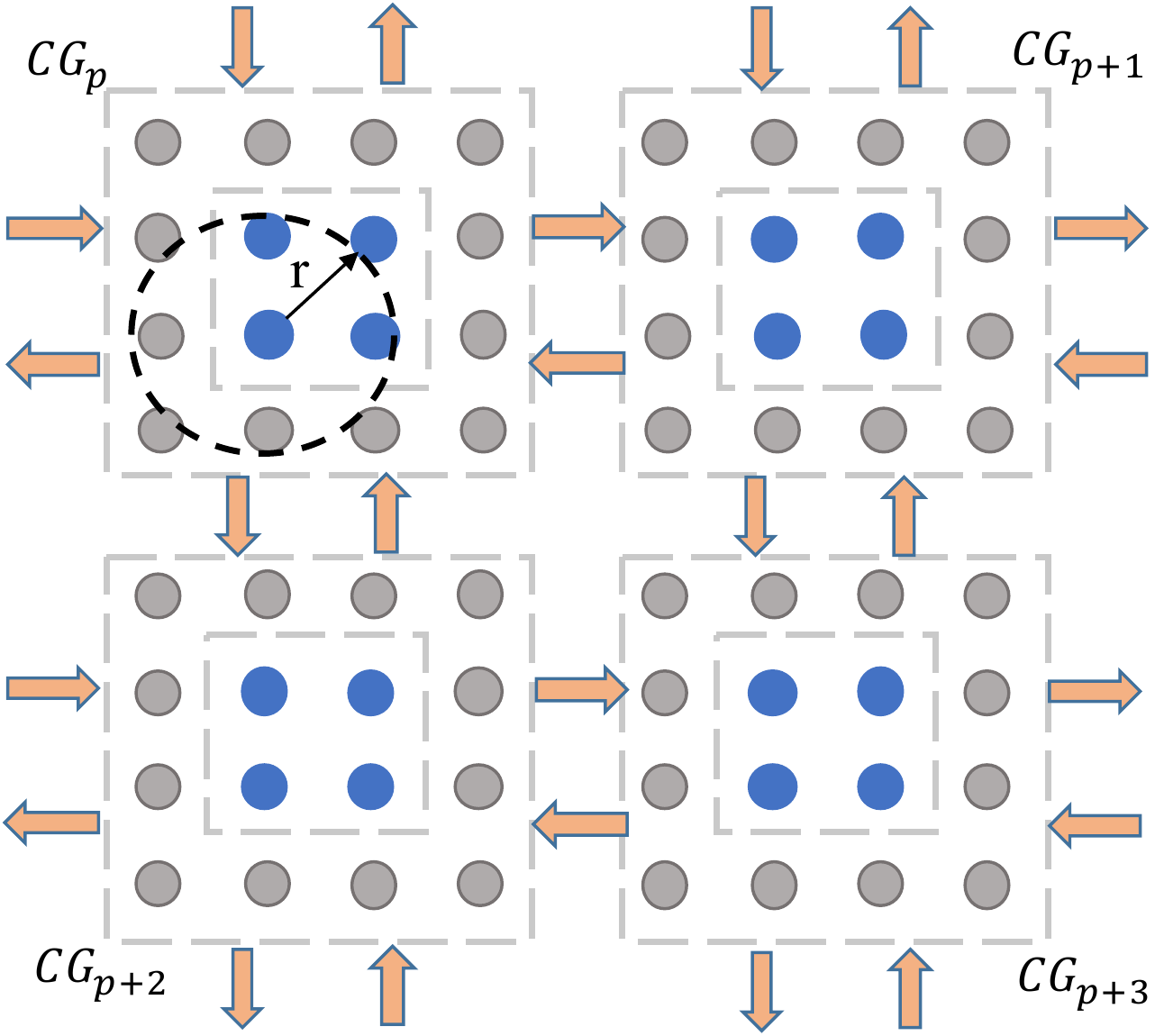}
\caption{The parallel processing among CGs using asynchronous MPI communication.}
\label{fig:fig10}
\end{figure}

For both CMP and CRS methods, the semblance calculation for a single CDP is the similar, however the calculation among CDPs is quite different. For the CMP method, there is no dependency among different CDPs. Therefore, for large-scale processing, we use the CDP as the granularity of a task. We divide the data into many partitions, and each MPE reads a separate data partition and processes the CDPs within the partition by assigning the CDP computation to the CPEs. After the processing of current CDP, the intermediate results are buffered before proceeding to the next CDP.
In the CRS method, each CDP calculates the two-dimensional coordinates of the middle point according to the coordinates of the source point and the receiving point. As shown in Figure~\ref{fig:fig10}, each CDP draws a circle based on its two-dimensional coordinates of the middle point. The data of all CDPs in this circle is collected by the central CDP to be processed. Therefore, for the CRS method we take the computation of the central CDP as the task granularity. We divide the coordinate grid by the coordinate of the middle point. The CDP at the boundary of the grid needs to obtain data from its neighbors, whereas the rest of the CDPs only need to obtain data within the grid.

Specifically, we pre-process the CDP data, during which the CDP data belonging to the same grid is written to the same data partition. Each MPE reads a data partition, aggregates the traces into corresponding CDPs, and then calculates the middle coordinates for each CDP. According to the radius $apm$ used in the CRS method, the MPE identifies the boundary of each grid, denoted as outer region, and leaves the rest as the inner region. As shown in Figure~\ref{fig:fig10}, the gray points belong to the outer region, whereas the blue points belong to the inner region. The MPE packs the CDP data of the outer region and sends it to the neighboring grids in the four directions, as well as receives the data from the neighboring grids. Since the calculation of the inner region does not require data from other grids, the MPE assigns the calculation of the inner region to the CPEs for parallelization. The calculation of the inner region and the data transfer of the outer region can be performed simultaneously bye using the MPI asynchronous communication. After the MPE asynchronously sends data through MPI, it calls CPEs to process the inner region in parallel. After the inner region is processed, the MPE checks whether the asynchronous communication finishes. After each grid receives the outer region data sent by its neighboring grids, each MPE proceeds to process the outer region of its own grid.

\section{Evaluation}
\label{sec:eval}

\subsection{Experimental Setup}
\label{subsec:setup}
In the experiments, we use the Sunway SW26010 processor for performance evaluation. For comparison, we use the-state-of-the-art implementations~\cite{gimenes2018evaluating} of the CMP and CRS methods running on 2 Intel E5-2680 V4 processors with 28 physical cores and Nvidia K40 GPU. We use the -O3 and -DNDEBUG options to compile the program. We also turn on the auto-vectorization during the compilation. We generate 8 diverse seismic datasets with detailed properties shown in Table~\ref{tab:realdata}. In general, the number of CDPs ($ncdps$) is proportional to the size of the dataset. Our synthesized datasets contain the number of CDPs ranging from 61,628 to 2,648,430. For a single CDP, the $fold$ describes the number of traces contained in a CDP, the $ns$ describes the number of samples in each trace, and the $dt$ determines the number of data per random data access. Since there are no public seismic datasets available, our datasets are synthesized with diverse properties that we believe to be representative. The performance metric used in the evaluation is $semblance\_trace/s$, which equals to the number of intersections produced by all semblance calculations divided by the total execution time.

In the field of seismic processing, single precision floating point is accurate enough to derive valid results~\cite{gimenes2018evaluating}. Hence, all evaluation results presented in this paper are in single precision floating point. In order to verify whether our approach affects the accuracy of CMP and CRS method, we provide the relative error of the results compared to the executions on CPU. In addition, we compare the relative error of our optimized parallel implementations on CPEs, the sequential implementations on MPE as well as the parallel implementations on GPU. Since the trend of the relative error is almost the same between CMP and CRS method, we only provide the relative error of CRS in Table~\ref{tab:relerr}. It is clear that the relative error of CRS running on Sunway is much smaller compared to running on GPU. In addition, the relative error of the parallel implementation on CPEs is almost the same compared to the sequential implementation on MPE. This demonstrates our approach hardly affects the accuracy of the CMP and CRS method.

\begin{table}
    \centering
    \caption{The detailed properties of the seismic datasets.}
   \begin{tabular}{c|c|c|c|c}
    \hline
     Seismic Dataset & fold & ns & dt & ncdps (large scale)\\
     \hline
    data1~ & 60 &550 & 220 &  2,648,430\\ \hline
    data2~ & 60 &550 & 240 &  2,648,430\\ \hline
    data3~ & 60 & 1,650 & 220 & 1,000,000\\ \hline
    data4~ & 60 & 1,650 &  240 &  1,000,000\\ \hline
    data5~ & 1,000 & 550 & 220 &202,500\\ \hline
    data6~ & 1,000 & 550 & 240 &202,500\\ \hline
    data7~ & 1,000 & 1,650 & 220 & 61,628\\ \hline
    data8~ & 1,000 & 1,650 & 240 & 61,628\\ \hline
  \end{tabular}
  \label{tab:realdata}
\end{table}


\begin{table}
    \centering
    \caption{The relative error of parallel implementation on GPU, sequential implementation on MPE and parallel implementation on CPEs of CRS method compared to CPU.}
   \begin{tabular}{c|c|c|c}
    \hline
     Seismic Dataset & GPU & $Seq_{MPE}$ & $Para_{CPEs}$ \\
     \hline
    data1~ & 5.88e-02 & 3.29e-05 & 3.29e-05\\ \hline
    data2~ & 7.44e-02  & 2.79e-05 &  2.79e-05 \\ \hline
    data3~ & 7.42e-02 & 1.22e-04 &  1.22e-04 \\ \hline
    data4~ & 7.76e-02 & 1.35e-04 &  1.35e-04 \\ \hline
    data5~ & 3.31e-02 & 1.43e-04 &  7.81e-03 \\ \hline
    data6~ & 4.40e-02 & 7.81e-03 &  7.81e-03 \\ \hline
    data7~ & 3.34e-02 & 6.7e-310 &  6.9e-310 \\ \hline
    data8~ & 5.24e-02 & 7.81e-03 &  7.81e-03 \\ \hline
  \end{tabular}
  \label{tab:relerr}
\end{table}

\subsection{Single Node Evaluation}
\label{subsec:singleeval}
\begin{figure*}
\centering
\includegraphics[scale=0.50]{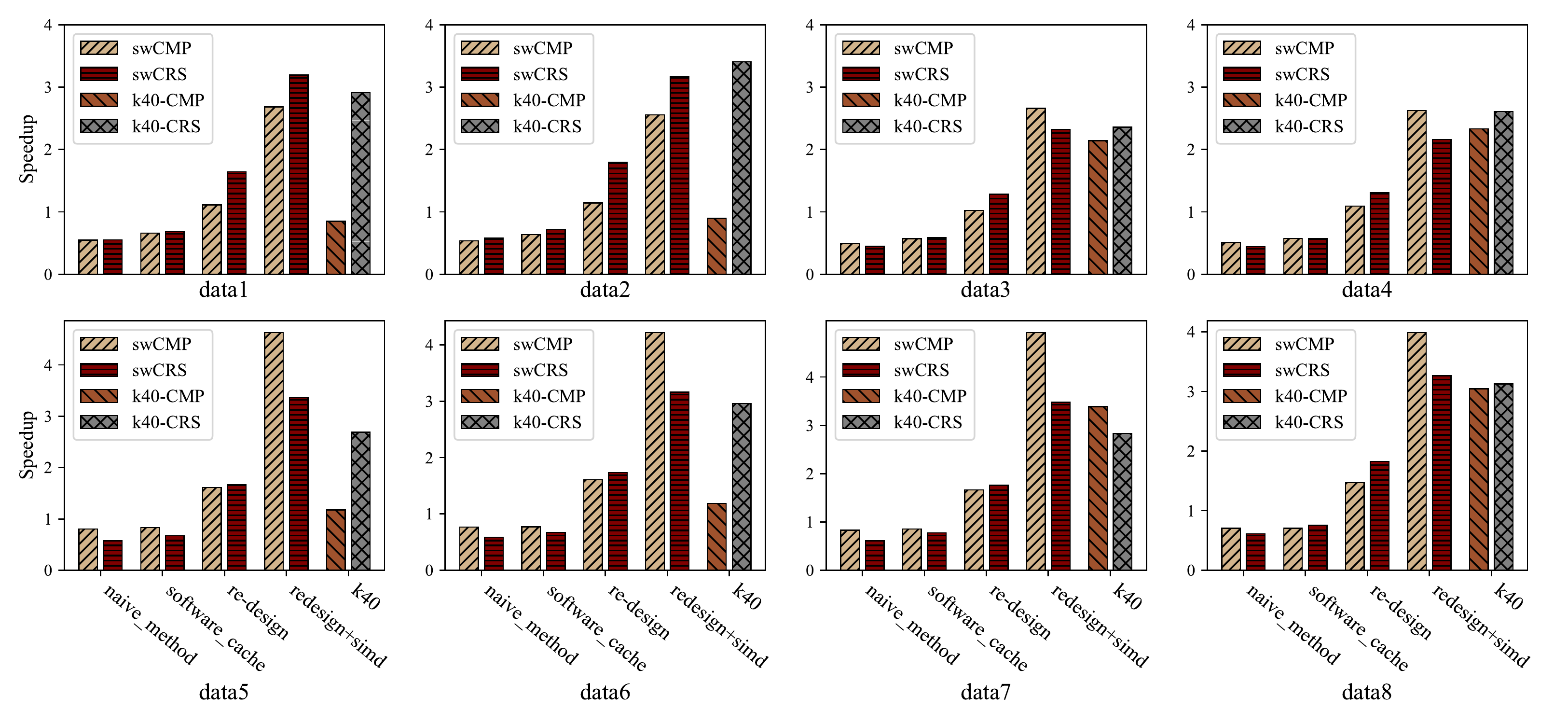}
\caption{The performance comparison of the CMP and CRS implementations on one Sunway processor, dual CPU and GPU K40. The speedup on the y axis is normalized to the baseline performance on CPU.}
\label{fig:fig12}
\end{figure*}

The performance comparison of the CMP and CRS implementations on one Sunway processor, dual CPU and GPU K40 is shown in Figure~\ref{fig:fig12}. 
We scale down the $ncdps$ of all datasets in Table~\ref{tab:realdata} to 8 in order to fit the resources on a single node across all architectures. The performance on dual CPU is chosen as the baseline. We also show the performance impact after applying our optimization techniques such as software cache, calculation re-design and vectorization (simd). As shown in Figure~\ref{fig:fig12}, the naive implementations on Sunway are limited by the memory bandwidth and cannot fully utilize the computation power of CPEs. It is also clear that our optimization techniques are quite effective to mitigate random memory accesses as well as exploit the vectorization for improving the performance of seismic processing on Sunway. After applying all our optimization techniques on Sunway, the CMP and CRS method achieves 3.50$\times$ and 3.01$\times$ speedup on average respectively across all datasets compared to the baseline. We notice that the CMP method achieves better performance in all eight datasets compared to GPU, whereas the CRS method is slightly worse than GPU on three datasets (\textit{data2}, \textit{data3} and \textit{data4}). This is mainly due to the limited memory bandwidth of Sunway processor (90.4GB/s), whereas the memory bandwidth of GPU K40 is higher by an order of magnitude (288GB/s).

\subsection{Roofline Model Analysis}
\label{subsec:roofline}
\begin{figure}
\centering
\includegraphics[scale=0.3]{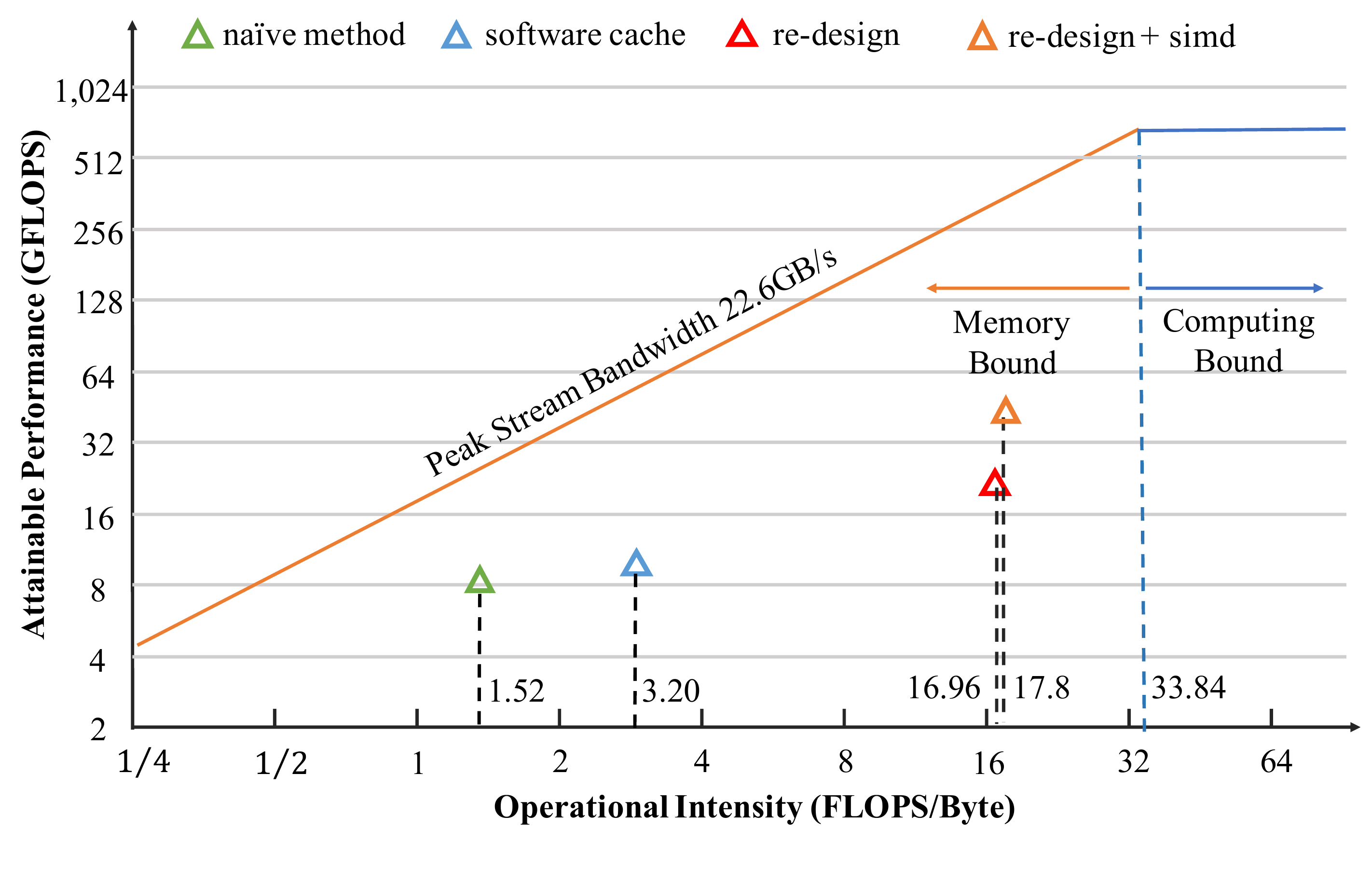}
\caption{The roofline model of the CRS implementation on Sunway.}
\label{fig:fig15}
\end{figure}

We use the roofline model analysis to better understand the performance impact of our proposed optimization techniques on Sunway. Due to the similar computation pattern between CRS and CMP, we only provide the roofline model analysis of CRS for conciseness. We analyze the performance results of CRS implementation on data1 dataset. Other evaluation results show the similar tendency. As shown in Figure~\ref{fig:fig15}, the operational intensity of the original program is 1.52 FLOPS/byte. In addition, the roofline model of a Sunway CG reveals that in order to fully utilize its performance, 33.84FLOPS calculations should be performed when accessing one byte data in memory. As shown in Figure~\ref{fig:fig15}, after applying our software cache, the operational intensity is doubled due to the data access by different intersections can be used by each other. In addition to the software cache, after re-designing the procedure of semblance calculation, the operational intensity $I$ can be derived using Equation~\ref{eq:intensity}. For a particular dataset, $w$ is a constant, the size of the tile is mainly determined by the size of the LDM, and $size\_get$ refers to the size of the data accessed by a DMA operation on a CPE. The more intersections processed by a single CPE at a time, the more data is overlapped and can be reused for latter calculation. The operational intensity after applying the calculation re-design increases to 16.96 FLOPS/byte. The roofline model analysis demonstrates our optimization techniques are effective to improve the performance of seismic processing on Sunway.

\begin{equation}
\label{eq:intensity}
I=\frac{tile \times (12+7\times w)}{size\_get\times 4}
\end{equation}


\subsection{Scalability}
\label{subsec:scalability}

\begin{figure*}
\centering
\includegraphics[scale=0.45]{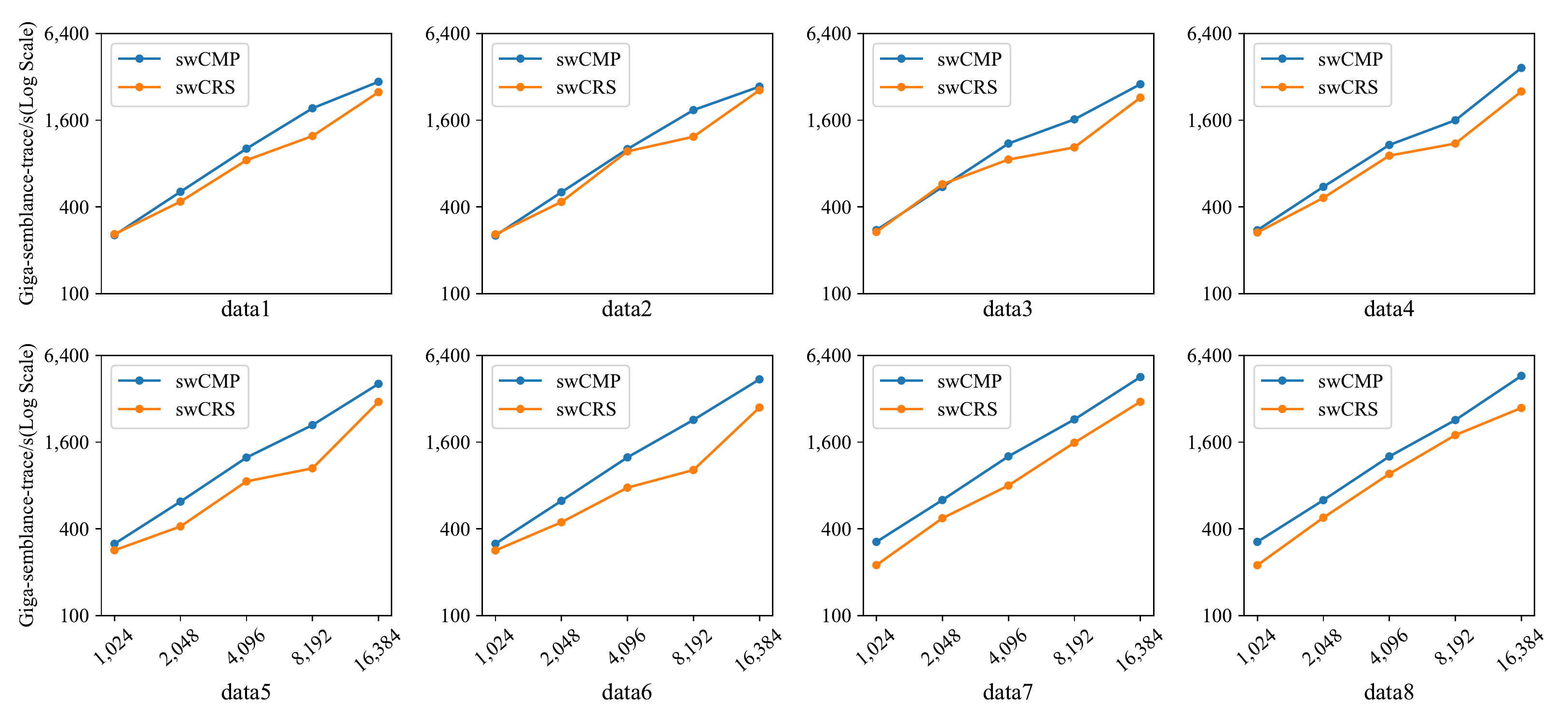}
\caption{Strong scalability for both CMP and CRS methods on Sunway. The x axis indicates the number of Sunway CGs, and the y axis indicates the performance in terms of giga-semblance-trace/sec (log scaled).}
\label{fig:fig13}
\end{figure*}

We evaluate both the strong and weak scalability of the CMP and CRS methods on Sunway. The performance is measured by $semblance\_trace/s$ of both methods excluding the I/O time. The size of the datasets ranges from 336GB to 418GB. For strong scalability, the number of CGs used for seismic computation scales from 1,024 to 16,384 with the input dataset unchanged. For weak scalability, when the number of CGs doubles, the size of the input dataset also doubles. We use the performance when running on 1,024 CGs as the baseline. The evaluation results for strong scalability is shown in Figure~\ref{fig:fig13}. Since the CMP method does not exchange data between processes, it maintains good scalability in general. Whereas the CRS method exchanges the boundary data between processes, its scalability is poorer than CMP method in all cases.

\begin{figure*}
\centering
\includegraphics[scale=0.45]{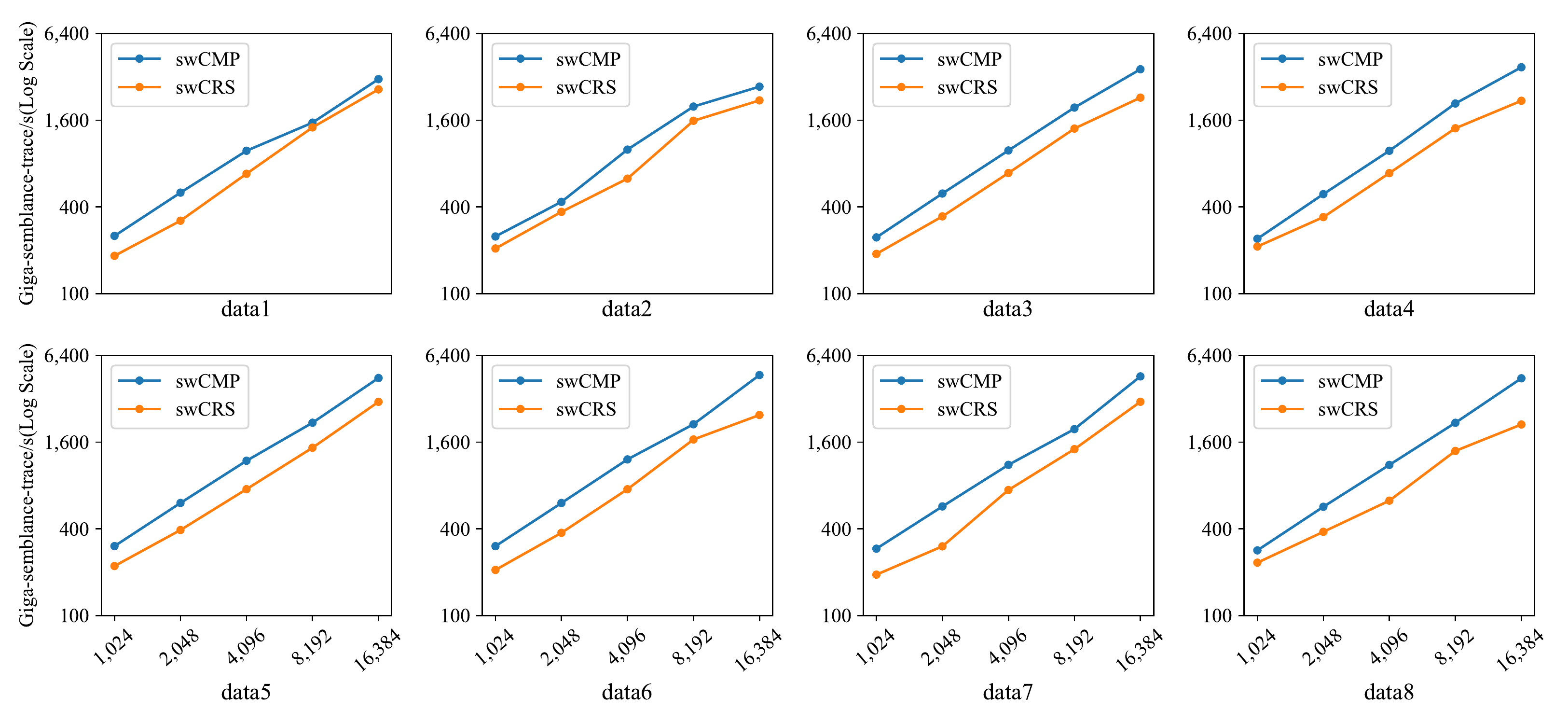}
\caption{Weak scalability for both CMP and CRS methods on Sunway. The x axis indicates the number of Sunway CGs, the y axis indicates the performance in terms of giga-semblance-trace/sec (log scaled).}
\label{fig:fig14}
\end{figure*}

Figure~\ref{fig:fig14} shows the evaluation results of weak scalability. We use 16,384 CGs to process the dataset with maximum size, and scale down the size of the dataset as the number of CGs decreases. Similar to the strong scalability experiments, the CMP method achieves better scalability compared to CRS in all cases. Note that each CG contains 65 Sunway cores, therefore the number of cores used in the experiments ranges from 66,560 ($1,024 \times 65$) to 1,064,960 ($16,384 \times 65$, more than one million Sunway cores!). The scalability results demonstrate our implementations of seismic processing are capable to run in large scale on Sunway TaihuLight supercomputer.

\subsection{Portability}
\label{subsec:portability}
Although the proposed optimization techniques are targeting the Sunway TaihuLight supercomputer, these techniques are also applicable on other systems~\cite{de2013clustered} that adopt the similar many-core cache-less architecture. Specifically, \textit{1)} the re-design of semblance calculation procedure increases the computing intensity of seismic processing significantly as shown in the roofline model analysis. This technique is effective to improve the performance of seismic processing on systems that lack L2 cache or with limited L2 cache; \textit{2)} the vectorization method improves the computation efficiency when processing the tiny data within short loops. This technique is necessary for seismic processing to exploit the powerful vector units with ever-increasing width on emerging architectures (e.g., AVX512 on Intel KNL).

\section{Related Work}
\label{sec:relatedwork}

\subsection{Performance Optimization of Seismic Processing}
\label{subsec:relatedseismic}

There has been a lot of work trying to improve the CMP method. Silva et al. ~\cite{da2016comparative} evaluate the performance of the CMP method on different platforms. The CMP method is implemented using the SYCL programming model and compared with the implementations using OpenCL and OpenMP. However, the evaluated platforms have high memory bandwidth, which dose not suffer the performance problem on Sunway due to the limited memory bandwidth. Zeng et al. ~\cite{zeng2016graphics} explore a different signal-to-noise ratio optimizer with the time-frequency domain-phase weighted stacking. They implement their method using the FFTW C library and the cuFFT CUDA library with significant performance improvement. However, these high performance CUDA libraries do not exist on the emerging architectures such as Sunway. Lawrens et al. ~\cite{lawrens2015implementation} analyze the characteristics of the CRS algorithm and applies the NUMA parallel computation scheme to optimize the CRS-Stack computation. Due to the unique architecture design of Sunway processor, existing optimization techniques of seismic processing are difficult to directly apply to Sunway. The above reasons motivate our work to re-design and optimize the CMP and CRS method to adapt to the Sunway architecture, so that they can fully exploit the massive computation power of Sunway TaihuLight.

\subsection{Performance Optimization on Sunway}
\label{subsec:relatedsunway}
A large number of applications have been optimized on Sunway Taihulight supercomputer. Duan et al. ~\cite{duan2018redesigning} have realized large-scale simulation of molecular dynamics, which fully exploits the architecture advantages of Sunway with the design of complex software cache. There are also research works devoted to optimize of the computation kernels on Sunway. For instance, Liu et al. ~\cite{liu2018towards} implement the efficient Sparse Matrix-Vector Multiplication (SpMV) on Sunway, which uses register communication to implement a complex communication mechanism, and thus achieves efficient mapping of SpMV algorithm to the hardware resources. Li et al. ~\cite{LiLYLQ18} implement an efficient multi-role based SpTRSV algorithm on Sunway. It leverages the unique register communication mechanism to address memory bandwidth limitations. Chen et al. ~\cite{chen2018simulating} re-design the earthquake simulation algorithm to reduce memory access costs tailored for the heterogeneous many-core architecture of Sunway.
All the above optimization works on Sunway have inspired our re-design and optimization techniques for the CMP and CRS method on Sunway. To the best of our knowledge, this paper is the first work to implement large-scale seismic data processing on the Sunway TaihuLight supercomputer with highly optimized CMP and CRS implementations targeting the Sunway architecture.
\section{Conclusion}
\label{sec:conclusion}
In this paper, we propose efficient implementations of seismic processing using both the CMP and CRS methods on the Sunway TaihuLight supercomputer for massively scaling. Specifically, we propose a software cache to alleviate the random memory accesses during the computation. We re-design the semblance calculation procedure to improve the bandwidth utilization by combining the search processes and buffering the intermediate results on each CPE. Moreover, we propose a vectorization method to improve the computation efficiency when processing tiny data within short loops. The experimental results show that our implementations of the CMP and CRS method on one Sunway processor achieve 3.50$\times$ and 3.01$\times$ speedup on average respectively than the-state-of-the-art implementations on CPU. Moreover, our approach is able to scale to more than one million Sunway cores with good scalability.

%

%
\bibliographystyle{ACM-Reference-Format}
\bibliography{sample-base}

%
\end{document}